# Fractional composition of large crystallite grains: a unique microstructural parameter to explain conduction behavior in single phase undoped microcrystalline silicon

Sanjay K. Ram[*,a], Satyendra Kumar[*,b] and P. Roca i Cabarrocas[!]

[*] *Department of Physics & Samtel Centre for Display Technologies, Indian Institute of Technology Kanpur, Kanpur-208016, India*

[!] *LPICM, UMR 7647 - CNRS - Ecole Polytechnique, 91128 Palaiseau Cedex, France*

We have studied the dark conductivity of a broad microstructural range of plasma deposited single phase undoped $\mu$c-Si:H films in a wide temperature range (15–450K) to identify the possible transport mechanisms and the interrelationship between film microstructure and electrical transport behavior. Different conduction behaviors seen in films with different microstructures are explained in the context of underlying transport mechanisms and microstructural features, for above and below room temperature measurements. Our microstructural studies have shown that different ranges of the percentage volume fraction of the constituent *large* crystallite grains ($F_{cl}$) of the $\mu$c-Si:H films correspond to characteristically different and specific microstructures, irrespective of deposition conditions and thicknesses. Our electrical transport studies demonstrate that each type of $\mu$c-Si:H material having a different range of $F_{cl}$ shows different electrical transport behaviors.

## 1. INTRODUCTION

The wide range of possible applications of microcrystalline silicon ($\mu$c-Si:H) is limited only by the inherent microstructural complexities and the consequently intricate electrical transport behavior of this material [1,2]. In general, electrical transport in $\mu$c-Si:H films shows an activated behavior, although different transport mechanisms may be observed depending on the film microstructure, doping and the range of measurement temperatures. Electrical transport in $\mu$c-Si:H at above room temperature has been explained using models that invoke potential barriers at grain boundaries (GB) [3] and percolation [4]. Electrical transport at low temperatures has been explained in terms of tunneling and hopping [5,6,7,8].

In the study of electrical transport properties in $\mu$c-Si:H, the time-honored concepts regarding the roles of changing crystallinity with film growth and amorphous phase in the system lose their relevance in the highly crystallized single phase $\mu$c-Si:H films having no distinguishable amorphous phase [2,9,10]. Because $\mu$c-Si:H is not a microstructurally defined unique material, a certain model that satisfactorily explains electronic transport for $\mu$c-Si:H material having a particular type of microstructure may not be applicable to another type of material. In this article we have presented the results of dark conductivity measurements conducted over a wide range of temperature (450–15 K) on a wide range of well-characterized highly crystallized single phase undoped $\mu$c-Si:H samples, addressing the current transport mechanisms in the context of the complex microstructure. Our study reveals that the fractional composition of constituent large crystallite grains correlates well to the film microstructure and morphology; and can be empirically used to classify the films into types having similar microstructure and electrical transport behavior.

## 2. EXPERIMENTAL DETAILS

Undoped $\mu$c-Si:H films were deposited at a low substrate temperature ($T_s \leq 200$°C) in a parallel-plate glow discharge plasma enhanced chemical vapor deposition system operating at a standard rf frequency of 13.56 MHz, using high purity $SiF_4$, Ar and $H_2$ as feed gases. We systematized our work by studying the influence of varying either gas flow ratio ($R = SiF_4/H_2$; $SiF_4$=1 sccm and $H_2$ dilution range 1-20 sccm) or $T_s$ (100-250°C) on the film microstructure, for different film thickness ($d \sim$ 50-1200 nm), thereby creating series of samples. The microstructural studies were carried out employing bifacial Raman scattering (RS), spectroscopic ellipsometry (SE), X-ray diffraction (XRD), and atomic force microscopy (AFM) [11]. The dark conductivity $\sigma_d(T)$ measurements were carried out on a large number of well annealed samples having different thicknesses, microstructures and morphological properties, using coplanar geometry in different experimental set-ups (above room temperature,

---

[a] Corresponding author. E-mail address: skram@iitk.ac.in, sanjayk.ram@gmail.com

[b] satyen@iitk.ac.in

300–450K; and low temperature, 300–15 K).

## 3. RESULTS AND DISCUSSIONS

The microstructural characterization studies of the material with RS and SE showed a high crystallinity of all the samples. The experimental SE data was fitted using Bruggeman effective medium approximation (BEMA) [12]. We have chosen published dielectric functions for low-pressure chemical vapor deposited polysilicon with large (pc-Si-l) and fine (pc-Si-f) grains as reference in the BEMA model [13]. For fitting of the experimental SE data, we have used a three layered structure optical model of films that consists of a bottom layer interfacing with substrate, a bulk layer and a top surface roughness layer.

SE data demonstrated a high crystalline volume fraction ($F_c$ >90%) in the bulk of the material from the initial stages of growth, with the rest being density deficit having no amorphous phase. The fractional composition of the films obtained from SE data revealed crystallite grains of two distinct sizes, which was corroborated by the deconvolution of RS profiles using a bimodal size distribution of large crystallite grains (LG ~70–80nm) and small crystallite grains (SG ~6–7nm) [11]. There is a significant variation in the percentage volume fraction of the constituent large ($F_{cl}$) and small grains ($F_{cf}$) with film growth, as evidenced by the analyses of RS, SE and XRD. With film growth the ratio $F_{cl} / F_{cf}$ rises and void fraction becomes negligible, resulting in a improved material density. The commencement and evolution of conglomeration of crystallites with film growth demonstrated by AFM corresponds to the appearance and increase in the LG fraction.

### 3.1. Electrical properties: above room temperature

Electrical conductivity was studied in context of deposition parameters ($R$, $d$ and $T_s$), as they are known to influence the film microstructure. At above room temperature, $\sigma_d(T)$ of all the $\mu$c-Si:H films follows Arrhenius type thermally activated behavior described by:

$$\sigma_d = \sigma_0 e^{-E_a/kT}, \qquad (1)$$

where $\sigma_0$ is known as conductivity pre-factor and $E_a$ as activation energy. With an increase in film thickness, $\sigma_d$ increases and $E_a$ decreases, irrespective of the deposition parameter values [10]. However, films having similar $d$, but deposited under different conditions show wide variation in the electrical transport behavior, due to the various effects of deposition conditions on the underlying film microstructure. For samples with similar $d$, inter-sample comparability regarding electrical transport behavior is possible only when the film microstructure is understood in the context of the fabrication history [10]. The correspondence between thickness or deposition parameters and film microstructure is not always systematic because the same microstructure can be achieved by the adjustment of one or more deposition parameters.

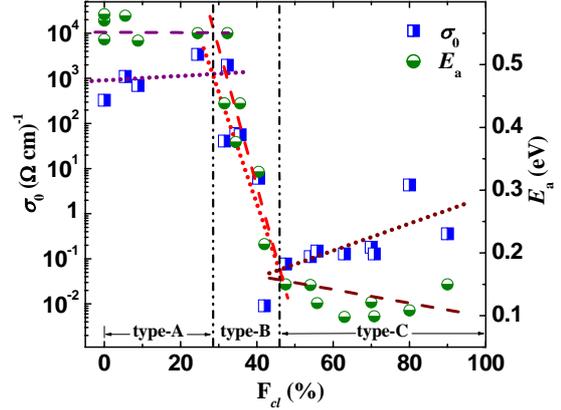

Fig. 1 Variation of $\sigma_0$ and $E_a$ with $F_{cl}$. Here left Y-axis depicts $\sigma_0$ and right Y-axis depicts $E_a$. The lines are shown to guide the eye.

Our earlier work has shown that $F_{cl}$ values correlate well to the stage of film growth and morphology, regardless of the deposition parameters [10]. Therefore, to explore the correspondence, if any, between the $F_{cl}$ values and electrical transport behavior, we have plotted the material parameters $\sigma_0$ and $E_a$ with the respective $F_{cl}$ values in Fig. 1. $E_a$ and $\sigma_0$ is valuable for understanding the mechanism and physics of electrical transport in the material. $\sigma_0$ is an indicator of the amount of statistical shift in Fermi level ($E_f$) and mobility edges. Here we see that three distinct and systematic trends of variation of the transport parameters are seen in different ranges of $F_{cl}$ (0–30%, 30–45% and > 45%). We have designated the films that fall into the three zones of this graph as *types – A, B and C* films. The variation of $\sigma_0$ and $E_a$ with film microstructure and $F_{cl}$ can be explained as follows. In the *type-A* films, the $F_{cl}$ increases from 0–35%. Here, the samples mainly contain SG ≤10 nm in size. Having a large number density of SG (with large number of SG boundaries) with a distribution of crystallite sizes leads to higher structural disorder even if there is no appreciable presence of an amorphous silicon tissue. Here $\sigma_0$ and $E_a$ are constant. Due to the predominant population of SG, the question of formation of potential barrier (i.e., transport through crystallites) does not arise because the large number of defect/trap sites compared to free electrons and small size of crystallites will result in a depletion width that is sufficiently large to become greater than the grain size, causing the entire grain to be depleted. Therefore, the transport in this type of material will be governed by the band tail transport.

In *type-B* films, $F_{cl}$ varies in the range of ~30- 40%. The electrical transport in this zone is strongly influenced by the formation of columnar boundaries, with changes in the transport routes. There is a sharp drop in $\sigma_0$ and $E_a$. The improvement in film microstructure leads to a delocalization of the tail states causing the $E_f$ to move towards the band edges, closer to the current path at conduction band (CB) edge ($E_c$). The statistical shift of $E_f$ depends on the temperature and the initial position of $E_f$, and when the $E_f$ is closer to any of the tail states and the tail states are steep, its statistical shift is rapid and marked.

The *type-C* films are constituted primarily of tightly packed crystallites, and have $F_{cl}$ >50%. The AFM studies of these samples had revealed that the conglomerate columns have an average width ~300–400nm. Our XRD results showed that preferred orientation is present in these films. The void fraction (density deficit) in these samples is negligible. Here, $\sigma_0$ shows a rising trend and the fall in $E_a$ is slowed down. A higher $F_{cl}$ and large size of columns result in less columnar boundary tissue with consequently lesser amount of associated defects. In addition, this boundary tissue forms a well established conducting network, which results in a rise in $\sigma_0$, and considering transport through this network, a band tail transport is mandatory [9]. The large columnar microstructure results in a long range ordering which is sufficient to delocalize an appreciable range of states in the tail state distribution. In addition, higher density of available free carriers and low value of defect density can cause a large increase in negatively charged dangling bond (DB) state density together with a decrease in positively charged DB states in the gap, which results in a lower density of states near the conduction band (CB) edge and can create a possibility of a steeper CB tail.

Thus, we see that in single-phase $\mu$c-Si:H material, in the absence of any correspondence with the total crystalline fraction (which is constant), or amorphous phase (which is absent), the above room temperature electrical transport behavior can be well correlated empirically to the film microstructure with the help of the percentage fraction of the constituent large crystallite grains.

### 3.2. Electrical properties: below room temperature

Now we come to the results of the low temperature electrical transport properties of our $\mu$c-Si:H samples belong to the three types described above. In Fig. 2(a) $\sigma_d(T)$ is plotted with reciprocal of $T$. The data reveals a continuously varying slope in $\sigma_d(T)$, which becomes almost independent of temperature for all samples at temperatures lower than certain values, depending on the film microstructure. This behavior indicates that the tunneling of carriers may be a dominant form of current conduction at lower temperatures.

According to Mott's variable range hopping (M-VRH) model for the three-dimensional case, if the density of states (DOS) is constant in a $kT$ energy range around $E_f$, the $\sigma_d(T)$ is expressed as: $\sigma_d = \sigma_0^* \exp(-T_M/T)^{1/4}$ where the terms $\sigma_0^*$ and $T_M$ are constants [14]. $T_M$ is related to the DOS at the Fermi energy ($N_f$) by expression: $T_M = C_M \alpha^3/(kN_f)$, where $k$ is Boltzmann's constant, $\alpha$ is the decay constant of localized wave function and $C_M$ is a constant. In Fig. 2(b), we have plotted $\sigma_d(T)$ data versus $T^{-1/4}$ and determined M-VRH constants. Here we see that *types-A* and *B* data fit better than the *type-C* data. According to Mott, the value of $C_M$ is $\approx 16$ for a constant DOS around the $E_f$. $\alpha R_{opt}(T) \geq 1$ is a prerequisite condition for M-VRH, which is found to be true in our case. However, Godet reported high values of $C_M$ (=310) for materials having an exponentially distributed DOS (G-VRH model) and set two conditions for this to be applicable [15]: First, a linear relationship should be present between $\ln \sigma_0^*$ and the slope $T_M^{1/4}$, which is evident in the inset of Fig. 2(b). The second condition states that the localization parameter (LP =$N_f \cdot \alpha^{-1}$) should be in the range of $10^{-5}$ -1, which also holds true for our case. Considering these conditions set out by Godet to discriminate between hopping near $E_f$ and hopping in bandtails, the G-VRH model seems to be

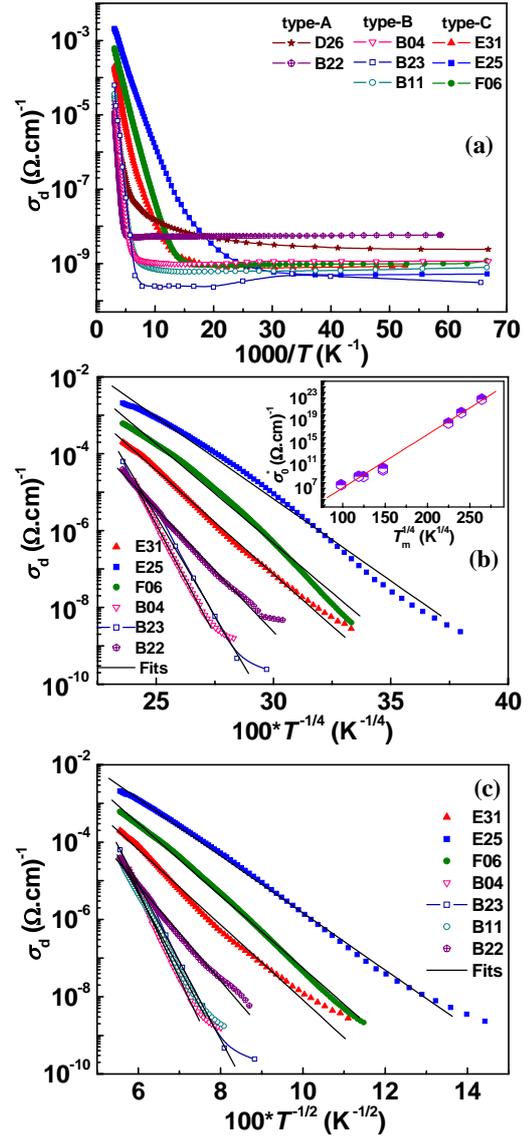

Fig. 2 (a) Temperature dependence of $\sigma_d(T)$ at low temperatures. The sample details are: *Type-A* samples: #B22 (170nm, $R$=1/10) and #D26 (410nm, $R$=1/5); *type-B*: #B11 (390nm, $R$=1/10), #B23 (590nm, $R$=1/10) and #B04 (950nm, $R$=1/10); *type-C*: #F06 (920nm, $R$=1/1), #E31 (1200nm, $R$=1/1) and #E25 (1025nm, $R$=1/5). $T_s$ of all the samples is 200°C. (b) $\sigma_d(T)$ data as a function of $T^{-1/4}$. The inset shows relationship between $\sigma_0^*$ and slope $T_M^{1/4}$. (c) $\sigma_d(T)$ data as a function of $T^{-1/2}$.

applicable to our observations. The reason behind this is probably that the improvement in film microstructure with film growth leads to delocalization of the tail states causing the $E_f$ to move towards the band edges. In contrast to the constant DOS at $E_f$ in amorphous silicon, the $E_f$ in μc-Si:H is actually somewhere in the vicinity of the exponential tail states. Therefore, in μc-Si:H system, both doped and intrinsic, the G-VRH mechanism is more applicable than the M-VRH.

A fitting of the same low temperature data to the conductivity expression, $\sigma_d \sim \exp(T_0/T)^{-1/2}$, is demonstrated in Fig. 2(c). Such a temperature dependence of carrier transport may arise either due to Efros-Shklovskii variable range hopping (ES-VRH) [16] or due to percolation-hopping (PH model) transport model as applicable to a composite system of granular metal dispersed in an insulator matrix proposed by Šimánek [17]. According to PH model, in a metal-insulator composite system, $T_0$ is given by: $T_0 = 16 \chi e^2 (s/d_g)^2 / k\varepsilon(½ + s/d_g)$, where $\chi$ is the rate of decay of the wave function in the insulator, $d_g$ is the average grain size, $s$ is the average width of the high resistive region between the neighboring grains, $\varepsilon$ is the dielectric constant of the insulator and $e$ is the electronic charge. The unreasonable material parameter values determined from $T_0$ using ES-VRH mechanism do not suggest it to be a possible transport mechanism in our case. In the heterogeneous structure of our μc-Si:H samples, the granular bulk (crystalline columnar island) conductivity is larger than the conductivity in the boundary regions containing disordered phase. The overall conductivity is governed by the resistive inter-granular region.

It is now desirable to see if the hopping parameters obtained from VRH mechanism ($T^{-1/4}$ dependence) and material properties deduced from $T^{-1/2}$ dependence are corroborative with the film microstructure by studying their variation with $F_{cl}$. The hopping parameters for G-VRH: $N_f$, optimum hopping distance, $R_{opt} = (9/8\pi\alpha N_f kT)^{1/4}$ and optimum hopping energy, $W_{opt} = 3/4\pi R^3 N_f$ are plotted in Fig. 3(a), (b) and (c) respectively. The values of $s/d_g$ obtained from $T_0$ ($T^{-1/2}$ dependence) of each sample using the value of $\chi \sim 1$ Å$^{-1}$, and $\varepsilon \approx 8$ for bulk-disordered region are plotted with $F_{cl}$ in Fig. 3(d). In Fig. 3(a), on proceeding from *type-A* to *type-B* material, we note a gradual fall in $N_f$ up to one order. The trend reverses and $N_f$ starts rising at about $F_{cl} \sim 40\%$ reaching a plateau in *type-C* above $F_{cl} \sim 50\%$. The observed variation of $N_f$ with film microstructure is explained by the changes occurring in the DOS with film growth. At the beginning of the growth, the DOS is high, and in spite of being fully crystalline, the film is mainly composed of SG, with a relatively high number of grain boundaries resulting in high defect densities. With film growth, the $F_{cl}$ rises, there is alteration in film morphology with the onset of conglomeration of grains, and the number of grain boundaries are reduced, resulting in a steep fall in the $N_f$ by an order of magnitude. Further, as we approach *type-C* material, $N_f$ rises, as the $E_f$ gets closer to CB edge, which is steeper compared to the CB edge in *type-A* and *type-B* materials.

In Fig. 3(b) and (c), the $R_{opt}$ and $W_{opt}$ values for *types*

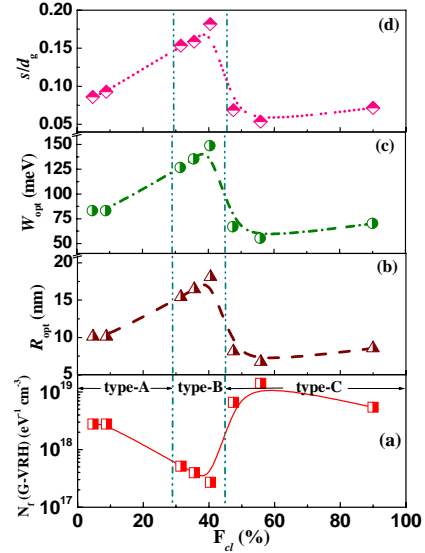

Fig. 3 Variation of G-VRH parameters, $N_f$, $R_{opt}$ and $W_{opt}$ with $F_{cl}$ are shown in parts (a), (b) and (c) respectively. The material property $s/d_g$ derived from Fig. 2(c) ($T^{-1/2}$ dependence) is plotted as a function of $F_{cl}$ in part (d).

*A* and *B* materials show a rising trend with rising $F_{cl}$. This variation of $R_{opt}$ and $W_{opt}$ are well correlated to the DOS variation with film growth, the lowering of DOS resulting in higher $R_{opt}$ and $W_{opt}$ values. As we reach *type-C* material, $R_{opt}$ and $W_{opt}$ start decreasing abruptly at $F_{cl} \sim 40$, reaching a low plateau level in *type-C* at $F_{cl} \sim 50$. For hopping model to be applicable here, $R_{opt}$ should be greater than or comparable to $d_g$, as the hopping distance has been physically correlated with the effective diameter of Si columns in Si nanostructures. When $R_{opt}$ approaches $d_g$, the hopping process corresponds to near neighbor tunneling, which should give rise to activated behavior of carriers. However, the $R_{opt}$ value in *type-C* is calculated to be 7-9 nm, which is less than the $d_g$ (~90–100nm), indicating that activated behavior should be evident in this temperature range, which contradicts our observations. Secondly, $\alpha^{-1}$ for highly crystalline polycrystalline material has been reported to be ~ 257Å [18]. Assuming such a value of $\alpha^{-1}$ for *type-C* material, ($\alpha \cdot R_{opt}$) comes out to be less than one, which precludes the basic assumption of $T^{-1/4}$ hopping mechanism.

Exploring the applicability of PH model in our material, in Fig. 3(d) we see that $s/d_g$ is low in the *type-A* material, which can be explained by the smaller boundaries of the densely packed small grains. The samples of *types A* and *B* in this study are all deposited under $R = 1/10$, where non-connecting conglomerate columns are formed having inverted pyramidal shapes. As a result, when the film grows in thickness, density deficit rises with significant void fraction between the columns/ grains. Thus, as we approach *type-B* material, these morphological changes occurring during film growth could result in a rise in the resistive boundary regions and the $s/d_g$ may rise. Nevertheless, the morphological studies of the *type-B* material suggest otherwise, and the onset of conglomeration should result in a lower $s/d_g$ than the derived values, because for any increase in $s$, there is also a significant increase in the $d_g$ of the conducting crystallite conglomerates. However, in *type-C* material, all the samples in this study are deposited under $R=$ 1/1 and 1/5. Here a least value of void fraction is seen, resulting in a highly dense material with tightly joined crystallites and straight conglomerate columns. The conglomeration is complete, and a sharp decline in $s/d_g$ signifies the low amount of boundaries compared to the larger width of conducting crystallite conglomerates. Our morphological studies have yielded $s \sim$ 5–7nm and $d_g \sim$ 90–100nm, which leads to $s/d_g \sim$ 0.05–0.07, a value closely matching the value calculated from this model.

Another corroborating evidence for $T^{-1/2}$ dependence in *type-C* material can be derived from the observed value of $T_0$ (Eq. 3), which is a material constant. $T_0$ has been reported to be $1\times10^4$–$3\times10^4$ K in highly crystallized nano or microcrystalline Si [7]. Value of $T_0$ has been calculated by theoretical method for nc-Si system by Rafiq *et al.*, yielding a value of $1.15\times10^4$ K [8]. $T_0$ values in our *type-C* material are $\sim 10^4$ K, while in *types A* and *B*, $T_0$ $\sim10^5$ K. Interestingly, $T^{-1/2}$ dependence of $\sigma_d(T)$ behavior reported for SiF$_4$ based plasma deposited $\mu$c-Si:H material is only seen in fully crystallized (100% crystalline volume fraction) material and not in mixed phase $\mu$c-Si:H system. Therefore, $T^{-1/2}$ dependence is more applicable in *type-C* material. Furthermore, a comparison between the two types of fitting ($T^{-1/2}$ and $T^{-1/4}$) using statistical parameters (like $\chi2$ or standard deviation) suggests $T^{-1/2}$ to be a better fit to the $\sigma_d(T)$ data of highly crystallized thick samples (*type-C*), but the same cannot be said for the thinner *type-A* samples conclusively.

Thus $T^{-1/2}$ dependence of $\sigma_d$ is seen in those $\mu$c-Si:H films whose microstructure consists of tightly packed large columnar grains without any distinguishable disordered phase (like *type-C* $\mu$c-Si:H), whereas the films having small grains with a fair amount of defect densities in the boundary regions (like *type-A* $\mu$c-Si:H) show a $T^{-1/4}$ relation.

## 4. CONCLUSIONS

We have studied the temperature dependent dark conductivity $\sigma_d(T)$ in the temperature range of 15 – 450 K of microstructurally well-characterized highly crystallized undoped $\mu$c-Si:H samples having a variety of microstructures. Our results show that the percentage fraction of constituent large crystallite grains can be used as an empirical parameter to correlate a wide range of microstructures to the electrical transport properties in this material. We have used this correlation to classify our samples into three types of material having different microstructures and electrical transport properties. The different underlying transport mechanisms for each temperature range have been discussed in the context of different film microstructures.

## ACKNOWLEDGEMENTS

One of the authors (SKR) gratefully acknowledges Dean of R & D, I.I.T. Kanpur, Samtel Centre for Display Technologies, I.I.T. Kanpur, Council of Scientific and Industrial Research, New Delhi and Department of Science and Technology, New Delhi, for providing financial support.

## REFERENCES

[1] R.W. Collins, A.S. Ferlauto, G.M. Ferreira, C. Chen, J. Koh, R.J. Koval, Y. Lee, J.M. Pearce, and C.R. Wronski, Sol. Energy Mat.& Sol. Cells 78 (2003) 143.
[2] J. Kocka, A. Fejfar, P. Fojtik, K. Luterova, I. Pelant, B. Rezek, H. Stuchlikova, J. Stuchlik, and V. Svrcek, Sol. Energy Mat.& Sol. Cells 66 (2001) 61, and references therein.
[3] P.G. Lecomber, G. Willeke, and W.E. Spear, J. Non-Cryst. Solids 59-60 (1983) 795.
[4] H. Overhof, M. Otte, M. Schmidtke, U. Backhausen, and R. Carius, J. Non-Cryst. Solids 227-230 (1998) 992.
[5] T. Weis, S. Brehme, P. Kanschat, W. Fuhs, R. Lipperheide, and U. Wille, J. Non-Cryst. Solids 299-302 (2002) 380.
[6] S. B. Concari and R. H. Buitrago, J. Non-Cryst. Solids 338-340 (2004) 331.
[7] M. Ambrico, L. Schiavulli, T. Ligonzo, G. Cicala, P. Capezzuto, and G. Bruno, Thin Solid Films 383 (2001) 200.
[8] M.A. Rafiq, Y. Tsuchiya, H. Mizuta, S. Oda, S. Uno, Z.A.K. Durrani, and W.I. Milne, J. Appl. Phys. 100 (2006) 014303, and references therein.
[9] D. Azulay, I. Balberg, V. Chu, J.P. Conde, and O. Millo, Phys. Rev. B 71 (2005) 113304, and references therein.
[10] S.K. Ram, P. Roca i Cabarrocas and S. Kumar, Thin Solid Films (2007) (in print).
[11] S.K. Ram, D. Deva, P. Roca i Cabarrocas and S. Kumar, Thin Solid Films (2007) (in print).
[12] S. Kumar, B. Drevillon and C. Godet, J. Appl. Phys. 60 (1986) 1542.
[13] G.E. Jellison Jr., M.F. Chisholm and S.M. Gorbatkin, Appl. Phys. Lett. 62 (1993) 3348.
[14] N.F. Mott and E.A. Davis, Electronic Processes in Non Crystalline Materials, 2nd ed., Oxford University Press, (1979).
[15] C. Godet, J. Non-Cryst. Solids 299 (2002) 333.
[16] A.L. Efros and B.I. Shklovskii, J. Phys. C 8 (1975) L49.
[17] E. Šimánek, Solid State Commun. 40 (1981) 1021.
[18] M. Thamilselvan, K. Premnazeer, D. Mangalaraj, and S. K. Narayandass, Physica B 337 (2003) 404.